\documentclass{aa}
\usepackage{natbib}
\usepackage{txfonts}

\def\gtorder{\mathrel{\raise.3ex\hbox{$>$}\mkern-14mu
             \lower0.6ex\hbox{$\sim$}}}
\def\ltorder{\mathrel{\raise.3ex\hbox{$<$}\mkern-14mu
             \lower0.6ex\hbox{$\sim$}}}

\def\Msun{M$_{\odot}$}

\begin{document}
\title{The Origins of the Substellar Companion to GQ Lup}
\author{John H. Debes \and Steinn Sigurdsson}

\institute{Department of Astronomy \& Astrophysics, Pennsylvania State
University, University Park, PA 16802}

\abstract{
The recently discovered substellar  
companion to GQ Lup possibly represents a direct 
test of current planet formation theories.
  We examine the possible formation scenarios for the companion to GQ Lup assuming it is a $\sim$2 M$_{Jup}$
object.  We determine that GQ Lup B most likely was scattered into a large,
eccentric orbit by an interaction with another planet in the inner system.
  If this is the case, several directly observable predictions 
can be made, including the presence of a more massive, secondary companion 
that could be detected through astrometry, radial velocity measurements, or 
scuplting in GQ Lup's circumstellar disk.
This scenario requires a highly eccentric orbit for the companion
 already detected. 
These predictions can be tested within the next decade or so.  Additionally,
we look at scenarios of formation if the companion is a brown dwarf.  One
possible formation scenario may involve an interaction between a brown dwarf
binary and GQ Lup.  We look for evidence of any brown dwarfs that have been
ejected from the GQ Lup system by searching the 2MASS all-sky survey.

}
\authorrunning{Debes \& Sigurdsson}
\titlerunning{Origins of GQ Lup B}

\maketitle

\section{Introduction}
The recent discovery of a substellar companion to GQ Lup has started
an exciting phase of extrasolar planet studies, where direct images of 
planetary companions can inform substellar spectral models as well as models
of planet formation.  The inferred age and distance to the companion, as well
as its 
absolute photometry, is consistent with a lower limit of  
 $\sim$ 2 M$_{Jup}$ \citep{neuhauser05}.  The models that were used to 
determine the lower limit are based on those of \citet{wuchterl03}.

GQ Lup itself is
a 0.7\Msun\ star with an age of $\sim$1 Myr at a distance of 140 pc within
the Lupus 1 cloud \citep{neuhauser05,tachihara96,wichmann98,knude98}.  Since 
GQ Lup is a young star, it still posseses an unresolved circumstellar disk that
nevertheless shows up as a strong mid- and far-infrared excess \citep{hughes94,weintraub90}.  GQ Lup 
resides in a relatively isolated star forming region within the Lupus complex
of clouds, a string of four dark clouds that holds several small young stellar
associations \citep{hughes94}.

The companion, or GQ Lup B, was observed at a projected separation of 0.7\arcsec\ from its host star and has had its common proper motion confirmed.  At a 
distance of 140~pc, the projected separation is 98~AU.  It
has absolute K$_{s}$ and L$^\prime$ magnitudes of  13.1 and 11.7, corresponding
to a luminosity of $\sim$10$^{-2}$ L$_{\odot}$ \citep{neuhauser05}.  Additionally, low
 resolution spectra of the companion identify it as a $\sim$L1 type object with a
T$_{eff}$ of $\sim$2000 K.  Based on GQ Lup B's spectral type, low specific gravity, and infrared photometry, \citet{neuhauser05} concluded that GQ Lup B is most likely a planetary object.

The large separation between GQ Lup and its companion
provides a unique test for the formation of planetary systems,
since this system does not resemble the Solar System or indeed any of the other
known extrasolar planetary systems.  GQ Lup B's origin can
provide useful insights into how giant planets form.  In this paper, 
we endeavor
to sketch out the possible origin of GQ Lup B and to determine observational
signatures that test whether this formation scenario is correct.  In Section
\ref{s1} we determine the most likely scenario for the formation of
GQ Lup B, while in Section \ref{s2} we determine the observational signatures
of this scenario.  Finally in Section \ref{s3} we present our conclusions.

\section{Planet Formation Scenarios}
\label{s1}

\subsection{In Situ Formation}
It is unlikely that GQ Lup B formed in its current location given its extremely
large projected separation from GQ Lup.  Core accretion is the generally
accepted method of forming a planet like Jupiter \citep{safronov69}.  Core accretion formation
 times scale 
roughly as t$_{form} \propto a^2$, where $a$ is the object's semi-major axis
\citep{pollack96,safronov69}. This also assumes a surface density of the
protoplanetary disk that decreases as R$^{-3/2}$.  If it takes $\sim$
1 Myr to form an object like Jupiter at 5~AU, it will take 400 times longer 
than the lifetime of GQ Lup to form an object at GQ Lup B's orbital separation.
Also, such formation requires many orbital periods.  Assuming a circular orbit
with a semi-major axis equal to the projected separation,
GQ Lup B has had $\sim$10$^3$ orbital periods in the entire lifetime of the
star, whereas Jupiter required roughly 10$^5$ orbital periods to form in the
core accretion scenario. 

Another possibility is that this planet formed by
direct collapse of the gaseous material in the disk \citep{cameron78,boss97}.  
Such  a collapse generally takes several hundred orbital periods and requires
a relatively large gaseous protoplanetary disk, since the disk
must be gravitationally unstable.  Instability is roughly based on marginal
Toomre Q values, where Q$=c_s \Omega/\pi G \sigma$ where $c_s$ is the sound
speed, $\Omega$ is the rotation rate of the gas, and $\sigma$ is the 
surface density of the disk.  Disks tend to be stable at close radii, and 
as the disk gets cooler, becomes unstable at greater distances. While this
is the case, detailed simulations of marginally stable disks cannot
produce planetary companions at $\sim$100~AU 
because non-axisymmetric structure in the form
of spiral waves transport mass interior to large separations before a planet
can form \citep{boss05}.  These simulations also show that the spiral structure often does not overlap as is the case at smaller separations which enhances
the formation of planet forming clumps.

Since in situ formation is unlikely for a planet at such a large orbital 
separation, some mechanism of migration must be invoked for a plausible origin
to GQ Lup B.  In the next sections we look at various ways of forming a Jovian
planet closer to the central star and then moving it to its final position.

\subsection{Stellar Encounters}
\label{s2.2}
A close encounter between a young star with a protoplanetary disk and another
star can have direct effects on the orbital parameters of planets and on the
morphology of the surrounding disk.  Such encounters have been used as an explanation for the ring-like structure seen in the $\beta$ Pictoris disk as well as
the truncation of the Solar System's Kuiper belt \citep{kalas00,larwood01,ida00}. 

 The close
passage of a star will give a gravitational kick to any planets in orbit and
potentially pump up their eccentricities.  If we assume that a $\sim$2~M$_{Jup}
$ object formed at a distance of $\sim$10~AU, we can infer the magnitude of 
the object's eccentricity based on its current location.  If such an object 
received a gravitational kick, its periastron would be roughly its primordial
 semi-major axis assuming it was initially in a circular or close to circular 
orbit.
After the kick the new orbit of the object would be highly eccentric, and its 
current position would correspond to apastron.  Ignoring projection effects, 
the planet's orbital eccentricity would be $\sim$0.8 and its orbital 
semi-major axis would be 54~AU.

A star's passage must be close to effect such a large eccentricity boost.  If
we assume the kick is impulsive, the star has to pass within two times the 
semi-major axis to produce an eccentricity $\sim$0.8.  One can calculate
how frequent such a passage would be.  We assume the cross section radius
to be $\sim$20~AU,
corresponding to a close approach capable of pumping the eccentricity to $\sim$0.8 and a relative velocity of $\sim$3 km s$^{-1}$.  
Given these parameters, one would need a stellar
number density of $\sim$10$^{6}$ pc$^{-3}$ to attain a rate of one encounter
per Myr for stars in the Lupus 1 cloud.
  This density is orders of magnitude larger than
the typical densities of T associations, which are loosely bound groupings of
stars that number in the hundreds.  Assuming a density of closer to 10$^3$~pc$^{-3}$, the probability of such a close encounter occurring in 1 Myr is 10$^{-3}$.  
  
\subsection{Orbital Migration}
Torques between a circumstellar disk and a young planet often cause
inward orbital migration of the planet towards its host star 
 \citep{nelson04}.  However,
in many stellar systems, the circumstellar disk is also photoevaporated over time,
either by the central star or by outside sources.  In these cases, the disk
is removed by a wind which corresponds to a time varying surface density
in the disk.  For mass loss rates an order of magnitude less than those observed in the Orion nebula, significant outward migration of planets can occur \citep{veras04}.

In the case of GQ Lup, this mechanism is not plausible for several 
reasons.  Firstly, the mechanism of mass loss in the case of GQ Lup 
requires external irradiating O or B stars not 
currently found in the rather isolated Lupus cloud.  Secondly, this mechanism
cannot export the planet to orbital separations of $\sim$100~AU within 1 Myr.
Finally, such a mechanism would require the presence of a disk extending
out to roughly
the same orbital radius, which would have a high mass loss rate but still 
remain to both form the planet and move it outwards from an initial birthplace
of $\sim$10~AU.  Another plausible mechanism for the
truncation of the disk could
 again come from a stellar encounter, this time with an encounter distance of a
few hundred AU.  Assuming a truncation event at $\sim$200~AU, the probability 
of that happening over 1 Myr in a 10$^3$ pc$^{-3}$ T association is 
$\sim$10$^{-2}$.  It is unclear if such a truncation would propel a planet out
to the very edge of the disk in a time shorter than 1 Myr.

\subsection{Planet-Planet Scattering}
The scenario of planet-planet scattering is the most convincing origin for
a widely separated planet.  We predict that in the
GQ Lup system its planetary companion
 was formed in the inner system and was scattered into an eccentric orbit
through interactions with at least one more massive planet \citep[see also][]{boss05}.  
Two jovian mass planets can become unstable to close approaches and can scatter
into configurations where one planet is kicked into a highly eccentric orbit
\citep{weidenschilling96,ford01}.  Moreover, this process can mirror the observed
eccentricity distribution of the known planetary systems discovered by
radial velocity surveys if 
the planets have unequal masses that follow an M$^{-1}$ mass distribution
 \citep{ford03}.  
 Planet-planet
 interactions can cause outward orbital migration, leaving planets in wide, 
eccentric orbits \citep{marzari02,veras04}.  Furthermore, evidence for such interactions
in currently known extrasolar planetary systems points to a common occurence
of interactions among protoplanetary systems \citep{ford05}.

Given the numerical results presented to date, one can estimate the current 
orbital parameters of the planets.  Such estimates will be useful for testing the ultimate origin of GQ 
Lup B by providing directly observable tests for this scenario.  Given that
the region for the formation of giant planets seems to be $\sim$5-10~AU, 
we will assume that GQ Lup B formed in this region and was kicked out to its
present orbit by a currently unseen inner companion that remained in this
general region.  Once again, we can estimate the outer planet's
 final eccentricity based on its current projected separation and this assumed initial semi-major axis and derive an $e\sim0.8-0.9$.  
Inspection of the various cumulative distribution
functions of \citet{ford01} and \citet{veras04} demonstrate that such large
eccentricities are possible, depending on the mass ratio of the planets to the
star and the ratio of the two planetary masses.  Furthermore, the percentage
of systems that result in a wide, eccentric orbit is on the order of a few 
percent.  If each star in the Lupus cloud has suffered one scattering event in
1 Myr,
then it is likely that one system in the cloud would have a widely separated
companion like GQ Lup B. 

The inner planet is likely a more massive planet than
GQ Lup B itself, given that lower mass objects are often pumped to large 
distances in two body encounters.  Given the M$^{-1}$ probability 
distribution of masses, and picking between 2-12 M$_{Jup}$, one would expect 
the second planet to have a 68\% chance of being between 2-7 M$_{Jup}$.  
Its orbital semi-major axis should be smaller but close
 to the initial semi-major axis within a factor of two.  Its eccentricity,
assuming it does not suffer from orbital migration through interaction with
a circumstellar disk, should be $>$ 0.1 \citep{ford01,veras04}.

Another possibility is that this is a system in the process of ejecting one 
of its planets.  Given the young age of the star, the first unstable interaction between the two objects would have occurred toward the latter stages of
the formation timescale.  Since this timescale is most likely on the order of 1 Myr, the first kick probably occurred within the last 10$^5$ yr.  The current
orbital timescale of GQ Lup is $\sim$500 yr assuming a semi-major axis of 
54~AU (corresponding to an orbit with $e$=0.8).  This corresponds to $\sim$200
orbital periods of the outer planet.  Since ejections of planets can happen 
slowly over several conjunctions between two planets, if one or both orbital
solutions for the companions to GQ Lup are known, one can predict the
likelihood of ejection.

Finally, this scenario is not limited to the interaction of two planets \citep{marzari02}.  A 
similar scenario would play out with multiple planets, though the
orbits of the other planets would be harder to predict.  

\section{Observational Implications of Each Scenario}
\label{s2}
While many of the proposed scenarios are implausible, each has their own
observational consequences, which in combination with plausibility
will allow a choice 
between the possible formation pathways.  In situ formation and outward 
migration through interaction with a disk will result in a circular orbit
for GQ Lup B.  Stellar interaction would leave GQ Lup B in a highly eccentric
orbit, as would planet-planet interactions.  Planet-planet interactions 
require at least one more planet that is more 
massive and in an orbit consistent with a recent interaction 
present in the inner system.  
Since planet-planet interactions require the presence of a second, more massive
planet in closer orbit around GQ Lup, the quickest direct test of all of these
scenarios is to study GQ Lup for radial velocity or
astrometric variations.  Given our 
possible range of masses and semi-major axes for the second companion, we will 
look at two benchmark cases which represent the best and the 
worst possible cases.  The best case would be a 7 M$_{Jup}$ object in a 2.5~AU
orbit, which we denote as Case 1, the worst case
 a 2 M$_{Jup}$ object in a 10~AU orbit, which we denote as Case 2. 

 For astrometric observations,
the increase in mass is offset by a decrease in orbital semi-major axis and one
expects $\sim$0.2 mas amplitudes for both cases in orbital motion, excellent 
for SIM or high precision differential astrometry from the ground \citep{ford03b,lane04}.  The big difference will be the timescale for these measurements, 
since the orbit of Case 1 will have a period of $\sim$5 yr, and Case 2's orbit
will have a period of $\sim$40 yr.  For radial velocity measurements, both objects should
be detectable within several years, either as a positive velocity trend or
a full orbit, depending on where the object is between Case 1 and Case 2.  For 
Case 1, we would expect a velocity amplitude (assuming circular orbits)
close to 160 m s$^{-1}$, and for Case 2 the amplitude would be closer to 22 m
s$^{-1}$.  \citet{neuhauser05} have monitored this object since 1999,
so part of the parameter space spanned by Case 1 and Case 2
will be directly tested within a few years.  Of concern is the fact that 
GQ Lup is a young star, and has a higher intrinsic radial velocity scatter.
The scatter in m/s can be be estimated based on the amount of photometric
variability a star has.  For GQ Lup, its variability would be estimated to
be $\sim$6.5$(0.4 \delta V)^{0.9}v\sin{i}$ m/s \citep{saar97}.  GQ Lup is variable,
with an amplitude of $\Delta V\sim0.34$, corresponding to a characteristic
scatter in radial velocity of $\sim$74$v \sin{i}$ m/s.  It may be difficult to 
detect a planetary companion with this intrinsic radial velocity scatter, though the orbital timescale of the companion is much larger than the timescale of
GQ Lup's variability.

There are also several secondary and more indirect ways of inferring the 
presence of a second companion close to the star.  An SED of GQ Lup's 
circumstellar disk could show the presence of a gap or hole caused by the more
massive second companion.  IRAS and 2MASS data have already been taken, and 
when combined with {\em Spitzer} photometry or spectroscopy, could show the
presence or absence of gaps or a clearing in the center of GQ Lup's 
circumstellar disk.  

\section{Brown Dwarf Formation Scenarios}
Since the upper mass limit of GQ Lup B would make it a brown dwarf, 
it is an interesting test of
brown dwarf formation as well and adds to the paucity of known
brown dwarf companions to stars.  Many of the postulated explanations for 
the formation of brown dwarfs are primarily interested in unbound objects
or those in brown dwarf-brown dwarf binaries.
The canonical example of brown dwarf companion formation is a kind of stalled 
 fragmentation where small cores form in the disks of larger primary
stars or in star forming filaments and are weaned from their accretionary clouds by being ejected from their natal source of material \citep{reipurth01}.
For this scenario to work, GQ Lup itself must retain the amount of material
it needs to become a 0.75\Msun\ star, while little material remains for its
companion at a separation of 100~AU.  Disk collapse with competitive
accretion seems to create brown dwarf-like companions, though the high mass 
ratio of GQ Lup to a brown dwarf is not seen in hydrodynamical simulations
made to date \citep{bate05}.  Furthermore, formation models of gravitationally
unstable disks suggest that brown dwarfs can form through disk fragmentation
at only orbital periods $\gtorder$2$\times$10$^4$ yr, which is much further than the
projected separation of GQ Lup B \citep{matzner05}.  
If this is true, the brown dwarf had
to migrate inwards on a timescale of $\ll$1 Myr.

Capture of a brown dwarf companion is about as
rare (if not moreso) than the frequency of close encounters with other stars
calculated in Section \ref{s2.2}, and therefore most likely not important.
Similarly, GQ Lup B would not be the remnant of a photoevaporated stellar core
since no high mass stars are present in this star forming region.  Another
idea for brown dwarf formation is the dissolution of brown dwarfs formed in
unstable orbits around binary stars \citep{jiang04}.  GQ Lup does not appear
to have a binary stellar companion, so this scenario is also ruled out.

The general argument against brown dwarf companions relies on the fact that
any low mass companion to a protostar 
will continue to accrete from a circumbinary disk until
its mass is comparable to the primary \citep{jiang04}.  This scenario
 requires a disk to remain
present, but in a small number of cases, the circumbinary disk could
be truncated through interactions with other stars in the star forming region.

For GQ Lup, this would require a truncation far in excess of 100~AU since it
is not a dense cluster, coupled with the possible limitation to disk 
fragmentation at large orbital separations.  This process should also form brown dwarf/brown dwarf+star triple systems that are loosely bound.  In a situation similar to that
described by \citet{jiang04}, the triple could be unstable on short timescales
and dissolve, leaving an ejected brown dwarf and a brown dwarf/star binary with
a smaller semi-major axis than where the brown dwarfs initially formed.
Given the low frequency of brown dwarf companions to stars this process need
not be terribly efficient, just efficient enough to produce the observed 
systems.   

In this case, a directly observable consequence occurs
in the form of another brown dwarf object within a distance $\sim\tau_{ej}v_{esc}$, where $\tau_{ej}$ is the timescale since the ejection of the the third
brown dwarf, and $v_{esc}$ is the escape velocity of the brown dwarf.  The
brown dwarf should be less massive than GQ Lup B.  The
timescale for ejection must be less than the age of the system, or 1 Myr, and
we assume a maximum $v_{esc}$ of 1 km s$^{-1}$ corresponding to an ejection
at an orbital separation of $\sim$800~AU, the maximum distance from
GQ Lup would be roughly 1~pc or 0.5$^\circ$ at a distance of 140~pc.  
Any brown dwarf within this 
distance would be an ejection candidate.  {\em Spitzer} 
IRAC mapping of this region would
be fairly straightforward, 
since a sensitivity of $\sim$1mJy would be required for 
each image based on the mid infrared photometry of GQ Lup B.  Since the field
of view for IRAC is 5\arcmin$\times$5\arcmin, a map of the entire region
surrounding GQ Lup B could be done for all 4 IRAC filters in roughly 1.5 hrs
time, assuming 0.6s exposures with 5 dither points at each mapping point.
At a S/N$\sim$ 5 sensitivity of 0.4 mJy, objects roughly ten times fainter than
GQ Lup B could be detected.

We also looked at all 2MASS sources within 0.5 degree of GQ Lup to determine
any brown dwarf candidates, using the criterion of the 2MASS brown dwarf 
searches, namely all objects with K$_s \leq$15.0, J-K$_s \geq$1.3 and no POSS 
counterparts, or R-K$_s$$\geq$5.5 \citep{kirkpatrick99,kirkpatrick00}.  Additionally we require that a 
candidate
 have K$_s >$13.1, the observed K$_s$ of GQ Lup B, thus ensuring that the 
candidate is
of a lower mass if at the same distance and age as GQ Lup.  
We find 56 candidates that would require further 
spectroscopic, Mid-IR, and proper motion follow-up that fit the near-IR photometry criteria and lack optical counterparts within 5\arcsec.
  Their positions and 2MASS magnitudes are listed.

\section{Conclusions}
\label{s3}
We have shown that the most probable scenarios for the formation of GQ Lup
B.  If it is a planet, it was most likely formed in the inner system and
ejected outwards.  Such a scenario will require the presence of a second planet in an orbit that will be 
between 2.5 AU and 10 AU, with significant eccentricity and an observational signature that may be detectable in the near future through radial velocity 
measurements or astrometry.  If GQ Lup B is a very massive planet or brown 
dwarf, it is less clear how it formed.  If it formed through an interaction
between a brown dwarf binary and GQ Lup, then a nearby brown dwarf should 
have proper motion consistent with an ejection event.

Finally, it is interesting to speculate on the implications the confirmation
of GQ Lup B's planetary status has for the frequency of planets
in wide orbits.  If GQ Lup B is a common occurence, we would expect many other
discoveries to have already been made.  However, we can get an idea of the
upper limit of this frequency by looking at a recent survey
 for substellar objects
in wide orbits \citep{mccarthy04}.  For objects with masses $\sim$5 M$_{Jup}$,
 this survey
has found 0/42 stars with planets at separations $>$75~AU or an 80\% probability
that 
less than 4\% of stars had planets at those separations.  Assuming that 
GQ Lup B is 2 M$_{Jup}$, the upper limit to the occurrence of such objects $>$ 75~AU
would then be 2.5 times greater assuming the M$^{-1}$ probability distribution
of radial velocity surveys holds for this population of planets.  However,
more detailed observational work would need to be done to better constrain this
frequency.

Since the uncertainty in GQ Lup B's mass is large and the upper limit for the
mass is not planetary, the possiblity remains that the object is a brown
dwarf.  Our results hold for any widely separated
planet discovered.  For example, the planets that are postulated to be present 
around Formalhut and HR 4796A would imply the presence of another planet
 that is most likely more massive \citep{kalas05,wyatt99}.  
This second planet should be in a closer orbit
that is eccentric as well.  Another example could be $\epsilon$ Eridani,
with a confirmed planet at a semi-major axis of $\sim$3.4~AU and an 0.1 M$_{Jup}$ companion postulated at a semi-major axis of 40~AU based on the sculpting of $\epsilon$ Eridani's
dust disk \citep{quillen02}.  

\bibliographystyle{aa}
\bibliography{gqlup}

\begin{table}
\begin{tabular}{ccccc}
\hline
\hline
    RA & Dec & J & H & K$_s$ \\
\hline
  
15 49 43.50 & -36 06 53.4 & 17.013 & 14.904 & 14.980 \\
15 49 50.89 & -35 34 26.0 &  17.159 & 16.100 & 14.959 \\    
15 49 24.68 & -35 35 12.5 &  16.413 & 15.472 & 14.952  \\   
15 49 29.17 & -35 25 32.6 &  15.734 & 14.693 & 14.264   \\ 
15 49 35.72 & -35 34 14.5 &  15.201 & 14.163 & 13.812  \\   
15 48 45.36 & -35 45 14.5 &  16.253 & 15.147 & 14.811  \\   
15 49 48.01 & -35 36 57.3 &  15.950 & 14.862 & 14.629  \\   
15 51 44.17 & -35 39 16.9 &  16.297 & 15.528 & 14.786  \\   
15 49 27.52 & -35 39 15.2 & 16.912 & 15.903 & 14.918  \\   
15 50 10.36 & -35 32 23.7 & 16.340 & 15.709 & 14.901  \\   
15 49 31.27 & -35 29 24.0 & 16.160 & 15.111 & 14.510  \\   
15 49 32.24 & -35 35 24.9 & 16.586 & 15.452 & 14.844   \\  
15 49 40.21 & -35 27 27.5 & 15.657 & 14.733 & 14.340   \\  
15 51 43.21 & -35 46 41.1 & 16.718 & 15.758 & 14.698    \\   
15 49 30.02 & -35 36 52.5 & 16.245 & 15.230 & 14.933   \\  
15 49 18.83 & -35 37 20.9 & 16.183 & 15.203 & 14.819   \\  
15 49 49.13 & -35 33 48.5 & 16.460 & 15.662 & 14.817   \\  
15 49 30.48 & -35 37 10.5 & 15.883 & 14.938 & 14.521   \\  
15 50 02.21 & -35 31 57.2 & 16.063 & 15.199 & 14.495   \\ 
15 49 54.42 & -35 37 00.4 & 16.295 & 15.362 & 14.992   \\ 
15 49 17.13 & -35 39 34.3 & 16.283 & 15.434 & 14.704   \\  
15 50 57.33 & -35 29 53.1 & 16.390 & 15.624 & 14.954    \\ 
15 49 42.41 & -35 36 48.9 & 15.734 & 14.566 & 14.230   \\  
15 51 38.26 & -35 46 31.5 & 16.627 & 15.807 & 14.984   \\  
15 49 33.06 & -35 28 56.8 & 16.554 & 15.478 & 14.968   \\  
15 51 27.21 & -35 31 49.5 & 16.797 & 15.803 & 14.955   \\  
15 51 42.11 & -35 34 20.4 & 16.254 & 15.558 & 14.934   \\  
15 49 39.25 & -35 26 32.9 & 15.875 & 14.912 & 14.465   \\  

 \hline
\end{tabular}
\caption{\label{tab}Table of Candidates}
\end{table}

\begin{table}
\begin{tabular}{ccccc}
\hline
\hline
RA & Dec & J & H & K$_s$ \\
\hline
15 49 52.62 & -35 35 26.5 & 16.014 & 14.792 & 14.600   \\  
15 49 52.19 & -35 38 57.3 & 16.682 & 15.457 & 14.926   \\  
15 49 31.87 & -35 25 23.8 & 14.830 & 13.716 & 13.360   \\  
15 49 51.73 & -35 38 20.9 & 16.326 & 15.285 & 14.673   \\  
15 50 04.74 & -35 23 35.9 & 16.347 & 15.511 & 14.949    \\ 
15 49 31.35 & -35 27 55.2 & 16.266 & 15.151 & 14.849    \\ 
15 49 11.25 & -35 52 23.6 & 16.109 & 14.361 & 14.699    \\ 
15 47 27.81 & -35 51 45.1 & 16.093 & 15.018 & 13.762    \\ 
15 47 16.69 & -35 29 03.0 & 16.089 & 14.995 & 14.630    \\ 
15 48 35.23 & -35 27 33.4 & 15.184 & 14.228 & 13.843    \\ 
15 47 53.12 & -35 42 51.1 & 16.249 & 15.323 & 14.869    \\ 
15 49 29.10 & -35 12 01.6 & 15.888 & 14.963 & 14.532    \\ 
15 48 36.80 & -35 29 31.7 & 15.961 & 15.242 & 14.619    \\ 
15 48 43.38 & -35 41 32.5 & 15.978 & 14.937 & 14.646    \\ 
15 47 44.27 & -35 23 29.7 & 15.703 & 14.881 & 14.332    \\ 
15 47 26.17 & -35 28 27.3 & 16.988 & 15.793 & 14.665    \\ 
15 47 45.02 & -35 26 59.3 & 15.639 & 14.720 & 14.267    \\ 
15 48 59.86 & -35 25 41.2 & 15.808 & 14.732 & 14.407    \\ 
15 48 12.92 & -35 16 07.9 & 15.864 & 14.789 & 14.412    \\ 
15 47 16.35 & -35 28 50.1 & 15.893 & 14.852 & 14.590    \\ 
15 48 40.06 & -35 39 24.3 & 15.749 & 14.878 & 14.445    \\ 
15 47 55.09 & -35 23 38.3 & 15.766 & 14.961 & 14.434    \\ 
15 47 58.83 & -35 26 07.7 & 16.621 & 15.703 & 14.965    \\ 
15 48 22.95 & -35 38 46.8 & 16.124 & 15.100 & 14.782     \\ 
15 48 38.02 & -35 34 55.7 & 16.254 & 15.422 & 14.914    \\ 
15 49 15.28 & -35 25 07.4 & 16.089 & 15.067 & 14.734    \\ 
15 47 55.64 & -35 21 24.2 & 16.463 & 15.359 & 14.997   \\  
15 47 37.47 & -35 32 41.6 & 16.261 & 15.233 & 14.907 \\
\hline
\end{tabular}
\caption{\label{tab2}Table of Candidates (cont'd.)}
\end{table}
\end{document}